\documentclass[%
	twocolumn,
	showpacs,
	showkeys,
	superscriptaddress
]{revtex4}

\usepackage[T1]{fontenc}
\usepackage[utf8]{inputenc}

\usepackage{amsmath}
\usepackage{amssymb}
\usepackage{amscd}		

\usepackage{graphicx}

\usepackage{mathptmx}

\newcommand{\beq}{\begin{equation}}
\newcommand{\eeq}{\end{equation}}
\newcommand{\ket}[1]{\vert#1\rangle} 
\newcommand{\bra}[1]{\langle#1\vert} 
 
\newcommand{\proj}[1]{\ket{#1}\bra{#1}}

\newcommand{\one}{\mathbb{I}}
\DeclareMathOperator{\Tr}{Tr}
\newcommand{\hil}{\mathcal{H}}

\newcommand{\abs}[1]{\left\vert#1\right\vert} 
\newcommand{\norm}[1]{\left\Vert#1\right\Vert}

\begin{document}

\title{Equation of motion for entanglement}

\author{Markus Tiersch}
\author{Fernando de Melo}
\affiliation{Physikalisches Institut der Albert--Ludwigs--Universit\"at,	Hermann--Herder--Str.~3, D--79104 Freiburg, Germany}
\author{Thomas Konrad}
\affiliation{Quantum Research Group, School of Physics, University of KwaZulu-Natal, Private Bag X54001, Durban 4000, South Africa}
\author{Andreas Buchleitner}
\affiliation{Physikalisches Institut der Albert--Ludwigs--Universit\"at,	Hermann--Herder--Str.~3, D--79104 Freiburg, Germany}

\date{\today}

\begin{abstract} 
We review an evolution equation for quantum entanglement for $2\times 2$ dimensional quantum systems, the smallest system that can exhibit entanglement, and extend it to higher dimensional systems. Furthermore, we provide statistical evidence for the equation's applicability to the experimentally relevant domain of weakly mixed states.
\end{abstract}

\maketitle

\section{Introduction}

In quantum mechanics, the dynamics  is introduced by virtue of the Schr\"odinger equation:
\beq
i \hbar \frac {\partial \ket{\psi(t)}}{\partial t} = H \ket{\psi(t)};
\label{SCeq}
\eeq
where $\ket{\psi (t)}$ represents the state of the system at time $t$, and $H$ is the Hamiltonian governing the evolution. For a time-independent Hamiltonian, solving the dynamics of a quantum system ``merely'' requires to diagonalize its Hamiltonian in order to arrive at the time evolution of a set of basis states -- the eigenbasis. Since the Schr\"odinger equation is linear with respect to the state, this immediately also provides the time evolution of any initial state, by expressing the latter in the eigenbasis of the Hamiltonian. Thus the knowledge of the Hamiltonian's eigenstates suffices to construct the evolution of \emph{any} other state as well as for every quantity which is a function of the state, e.g.\ fidelities, overlaps, and expectation values of observables.
However, for large quantum systems, diagonalization of the Hamiltonian quickly turns into a tedious venture.
It is thus mandatory to directly construct an evolution equation for the quantities of interest,  which comprises only the minimal set of information needed.

Entanglement, arguably the most pronounced signature of quantum mechanics, is an even more extreme example of the intractability mentioned above. Entanglement is a non-linear function of the state from its very definition. A state $\rho$ of a quantum system that is identified with the Hilbert space $\hil_A\otimes \hil_B$ is said to be entangled, if and only if it cannot be written as a separable state, i.e.\ in the form~\cite{werner}:
\beq
\rho=\sum_i p_i \rho^A_i\otimes\rho^B_i,
\eeq
where $p_i > 0$ and $\sum_i p_i=1$.  Therefore, in order to quantify the amount of entanglement inscribed in a state, one must search for a decomposition of $\rho$ which is the ``closest'' to a state with the above form -- where  the definition of ``close'' depends on the  entanglement measure being used. One possibility is the convex-roof construction~\cite{Uhlmann:2000}, by which the entanglement measure $E$ defined over the pure states is extended to mixed states $\rho = \sum_i p_i \ket{\psi_i}\bra{\psi_i}$ by:
\beq
E(\rho) = \inf_{\{p_i,\ket{\psi_i}\}} \sum_i p_i E(\ket{\psi_i});
\eeq
where the infimum is performed over \emph{all} possible decompositions of $\rho$ (see Ref.~\cite{EntDist} for other possible definitions). Two important characteristics of this expression must be noticed: {\it i}) $E$ is in general a non-linear function of the state since it is convex: $E \big(\lambda \rho_1 +(1-\lambda) \rho_2\big)\le  \lambda E(\rho_1) +(1-\lambda) E(\rho_2)$, and {\it ii}) it is hard to be evaluated due to the infinitely many decompositions. In fact, it was shown in~\cite{gurvits} that in general to decide if a given state is entangled or not is a NP-hard problem.

\begin{figure}
\begin{center}
	\includegraphics[width=\linewidth]{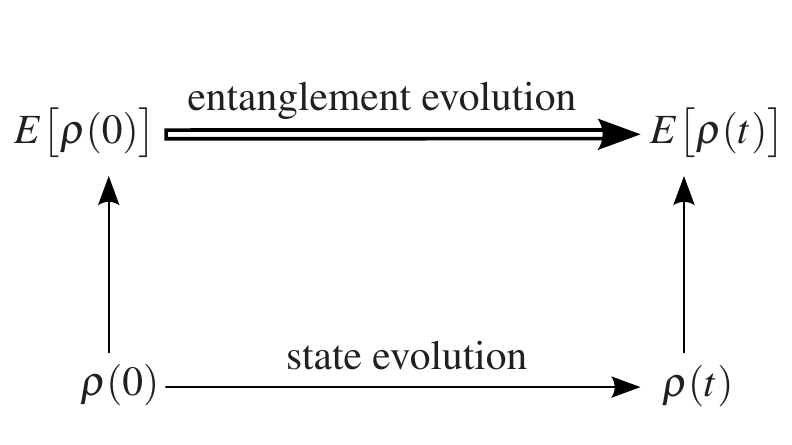}
\end{center}
\caption{
  From state evolution to entanglement dynamics. A direct equation of motion for entanglement, which uses only the minimal necessary information about the dynamics and initial state is desirable.
}
\label{arrows}
\end{figure}

The difficulty to determine the amount of entanglement is magnified when the quantum system undergoes some dynamics. Up to recently~\cite{Konrad:2008,Tiersch:2008} this required to follow the evolution of the state and subsequently, from the knowledge of all its parameters, calculate its entanglement (see Fig.~\ref{arrows}).  Experimentally this implies a full state tomography for each point in time, and for each initial condition. This is specially demanding in realistic  situations where quantum systems interact with uncontrolled or unobserved degrees of freedom, usually lumped together under the term \emph{environment}. The dynamics due to a coupling to an environment usually degrades the quantum features of the considered quantum system,  and it  is often exploited to explain the classical behaviour of the world despite it being quantum on the atomic level~\cite{Zurek:2002}. Foremost, in situations where true quantum effects such as entanglement provide a resource, e.g.\ for teleportation and various other quantum information and communication tasks~\cite{nielsen}, it is of primary interest to know how such resource decays while being processed or stored.

On the following pages we develop a deterministic equation for the evolution of quantum entanglement. This relation contains  only a single quantity which, independently of the initial state, describes how a state's entanglement evolves. This single quantity constitutes a benchmark for given open system and decoherence dynamics.

We start in section~\ref{basics} by reviewing some basic concepts and formulae concerning the entanglement measure of our choice. After that, in section~\ref{law},  we present an equation of motion for the entanglement of an initially pure state of two two-level systems when either one of them undergoes some evolution (as described by a completely positive map); extensions of this setup are also introduced. Finally,  in section~\ref{lawG}, we comment on the generalization of our framework to higher dimensional systems.

\section{Basics}
\label{basics}
\subsection{Concurrence}

We choose concurrence as a quantifier of entanglement~\cite{Wootters:1998} because of its convenient algebraic properties. For a pure state $\ket{\chi}$ of two finite-dimensional quantum systems, say with respective Hilbert spaces $\hil_A$ and $\hil_B$, it is
\beq \label{eq:concurrencePure}
C(\ket{\chi}) = \sqrt{2 \left(1-\Tr \rho_{A/B}^2 \right)},
\eeq
where $\rho_{A/B} = \Tr_{B/A} \proj{\chi}$ is the reduced state of either subsystem after tracing out the other. The compound quantum system of the smallest possible dimension which can exhibit entanglement is composed of two subsystems with two levels each. The corresponding pure state, when expressed in an arbitrary product basis of the respective subsystems, $\ket{\chi} = \sum_{i,j=0}^1 \chi_{ij} \ket{e_i}\ket{f_j}$, yields the expression for concurrence by means of a determinant of its coefficient matrix $\chi$:
\beq \label{eq:concDet}
C(\ket{\chi}) = 2 \abs{\det \chi}.
\eeq
When the quantum system undergoes open system dynamics and couples to its environment, the quantum state becomes mixed in general. The resulting mixed state $\rho$ is described by a probabilistic mixture of an ensemble of pure states, i.e., $\rho=\sum_i p_i \proj{\psi_i}$ with corresponding probabilities $p_i$ that sum to one. As previously stated, this measure can be extended over the mixed states by virtue of the convex roof construction~\cite{Bennett:1996b,Uhlmann:2000}:
\beq \label{eq:convexRoof}
C(\rho) = \inf_{\{p_i,\ket{\psi_i}\}} \sum_i p_i C(\ket{\psi_i}).
\eeq
In particular for states of high dimensional systems this optimization is difficult to perform. In fact, analytical expressions are only known for states of high symmetry~\cite{Vollbrecht:2001,Rungta:2003} and for systems of dimension $2\times 2$. In the latter case, the minimum is obtained by~\cite{Wootters:1998}
\beq
C(\rho) = \max \left\{ 0, \sqrt{\lambda_1} - \sqrt{\lambda_2} - \sqrt{\lambda_3} - \sqrt{\lambda_4} \right\},
\eeq
where $\lambda_i$ are the eigenvalues of the matrix $\rho (\sigma_y\otimes\sigma_y) \rho^* (\sigma_y\otimes\sigma_y)$ with $\lambda_1$ denoting the largest among them, and the complex conjugation is taken with respect to the product basis of eigenvectors of $\sigma_z$. Although this analytical expression for mixed states enables and encourages the treatment of  $2\times 2$ systems under exemplary cases of open system dynamics, we rather focus on the algebraic property introduced in \eqref{eq:concDet} to derive our results and build intuition.

\subsection{Entanglement under filtering operations}

In preparation of what follows, we briefly review how entanglement changes under filering operations~\cite{Gisin:1996,Bennett:1996a}. Such operations can be understood as the result of a measurement in a higher dimensional space. We consider a quantum system composed of two subsystems with state $\ket{\psi}=\sum_{i,j} \psi_{ij} \ket{i}\ket{j}$ and apply a filtering operation $M=\sum_{k,l} M_{kl} \ket{k}\bra{l}$ to one of the subsystems. After renormalization the state becomes
\beq
\ket{\psi^\prime}
= \frac{M \otimes \one \ket{\psi}}{\norm{M \otimes \one \ket{\psi}}} = \sum_{k,j} \psi^\prime_{kj} \ket{k}\ket{j},
\eeq
where the new matrix elements with respect to the product basis are $\psi^\prime_{kj} \propto \sum_i M_{ki} \psi_{ij}$, apart from the normalization factor, i.e.\ the product of the matrices: $\psi^\prime \propto M\psi$. Since for a matrix product the determinant fulfills $\det (M\psi)  =\det M \det \psi$, the concurrence~\eqref{eq:concDet} of the pure state after a filtering operation changes to
\beq
C(\ket{\psi^\prime}) = \frac{ \abs{\det M} }{ \norm{M \otimes \one \ket{\psi}}^2 } \, C(\ket{\psi}).
\eeq
If $M$ is invertible, e.g.\ it is not a projective measurement, the entanglement is not completely destroyed but merely rescaled by the determinant of $M$. Thus, for suitable combinations of filtering operations $M$ and states $\ket{\psi}$ entanglement can probabilistically increase.

Now, the treatment for pure states allows us to extend the above relation to mixed states. A filtered mixed state is
\beq
\rho^\prime= \frac{ (M\otimes\one)\rho(M\otimes\one)^\dag }{ \Tr[(M\otimes\one)\rho(M\otimes\one)^\dag] }
\equiv \frac{(\mathcal{M}\otimes\one)\rho}{ \Tr[(\mathcal{M}\otimes\one)\rho] } \;,
\eeq
where we abbreviated the action of $M$ onto $\rho$ as the action of a map $\mathcal{M}$.
Therefore, the  pure states of the decompositions of $\rho$ and its  ``filtered'' version $\rho^\prime$ are only related by  the filtering operation. This means that if the infimum in \eqref{eq:convexRoof} for $\rho$ is attained for a mixture of pure states $\{\ket{\psi_i}\}$ with respective probabilities $p_i$, then the optimal decomposition for $\rho^\prime$ is realized by $\{\ket{\psi_i^\prime}\}$ which are the filtered versions of the $\{\ket{\psi_i}\}$ with the same probabilities. Therefore we obtain for the change of concurrence of a mixed state under a filtering operation~\cite{Verstraete:2001}:
\beq \label{eq:filterMixed}
C(\rho^\prime) = \frac{ \abs{\det M} }{ \Tr[(\mathcal{M}\otimes\one)\rho] } \, C(\rho).
\eeq

\section{Evolution equations for concurrence}
\label{law}

In this work we model the effect of open system dynamics and in particular of decoherence on the quantum state by means of dynamical maps. This provides the required freedom to treat general open system dynamics without having to assume a specific interaction. Such maps act on density operators and mediate the evolution that the state undergoes during a certain time interval~\cite{BreuerPetruccione,nielsen}:
\beq
\rho(t) = \Lambda \big[ \rho(0) \big].
\eeq
The map $\Lambda$ is linear, preserves the trace, and maps quantum states to quantum states. In order to describe the dynamics of systems which are part of a larger system correctly, $\Lambda$ also needs to be completely positive. Any such map can be written as $\rho(0) \mapsto \sum_i K_i\, \rho(0)\, K^\dagger_i$, where $K_i$ are called Kraus operators and, due to the trace-preserving constraint, fulfill $\sum_i K^\dagger_i K_i = \one$. This dynamical map is usually termed Kraus or operator-sum representation~\cite{nielsen}.

We begin the treatment of the concurrence evolution under such dynamics with the simplest possible case, i.e.\ a $2\times 2$ system that starts in an initially pure state $\ket{\chi}$ and of which only one subsystem undergoes open system dynamics. A scenario that approximates this setup is a laboratory that prepares these two quantum systems in an entangled state, and subsequently sends one of the systems to a remote laboratory, for instance through a glass fiber in case of photons. During this process the transmitted photon suffers the detrimental influence of its environment, designated by $\Lambda$, whereas the other quantum system remains well protected in the lab. The system's final state thereafter is given by
\beq \label{eq:initialSetup}
\rho = (\one\otimes\Lambda) \proj{\chi}.
\eeq
We aim for a relation between the final and initial amount of entanglement, $C(\rho)$ and $C(\ket{\chi})$, respectively, in terms of $\Lambda$.

In order to derive such an evolution equation, we reformulate the initial setup~\eqref{eq:initialSetup} equivalently into its dual setup, i.e. we want to exchange the role of states and channels. In the first step, we express the initial state $\ket{\chi}=\sum_{i,j} \chi_{ij} \ket{i}\ket{j}$ as the result of a filtering operation $M$ onto the first subsystem of a maximally entangled state:
\beq
\ket{\chi} = M\otimes\one \ket{\phi^+},
\eeq
where we choose $\ket{\phi^+}=(\ket{0}\ket{0}+\ket{1}\ket{1})/\sqrt{2}$ as an exemplary maximally entangled state.
We thus define \linebreak $M=\sqrt{2}\sum_{k,l} \chi_{kl} \ket{k}\bra{l}$, which already includes  the normalization factor. Therefore, the same final state of the original setup \eqref{eq:initialSetup} can also be prepared via $(\mathcal{M}\otimes\Lambda)\proj{\phi^+}$. Since the filtering $\mathcal{M}$ and the open system dynamics $\Lambda$ act on different subsystems, their order may be interchanged. Applying $\Lambda$ first, we arrive at the dual setup:
\beq \label{eq:dualSetup}
\rho = (\mathcal{M}\otimes\one) \rho_\Lambda,
\eeq
where $\rho_\Lambda = (\one\otimes\Lambda)\proj{\phi^+}$ is a mixed state in general and contains all the parameters of the dynamics $\Lambda$. We have thus interchanged the roles of initial state and dynamics by moving all the parameters of the initial state $\ket{\chi}$ into a map $\mathcal{M}$, and the parameters of the dynamics $\Lambda$ into the state $\rho_\Lambda$. This is the essence of the Choi-Jamio\l{}kowski isomorphism~\cite{Choi:1975,Jamiolkowski:1972} from which the state $\rho_\Lambda$ is shown to be isomorphic to the map $\Lambda$.

We exploit this equivalence and apply the filering result \eqref{eq:filterMixed} to the dual setup \eqref{eq:dualSetup}. The resulting determinant of the filtering~$M$ equals the concurrence of the initial state $\ket{\chi}$ and yields the first result:
\beq \label{eq:factPure}
C\big[ (\one\otimes\Lambda) \proj{\chi} \big] =   C(\ket{\chi})\; C\big[ (\one\otimes\Lambda) \proj{\phi^+} \big].
\eeq
The  concurrence of all pure states evolves exactly as  that of a maximally entangled state, and is merely rescaled by the initial amount of entanglement. This relation clearly separates the initial condition from the evolution, which may be any dynamical process. The maximally entangled state thus serves as a benchmark according to which the entanglement of all other pure states evolve. This means that for a given open system dynamics $\Lambda$, there is qualitatively only one type of entanglement evolution. Also note that the particular kind of the maximally entangled state does not matter, since any other maximally entangled state, e.g.\ a singlet state, may be prepared from $\ket{\phi^+}$ by a single unitary operation on the ``left'' subsystem -- which changes $\rho_\Lambda$ but leaves its entanglement invariant. The relation holds also for more general $\Lambda$ than we initially assumed. Maps which do not perserve the trace, $\sum_i K^\dagger_i K_i < \one$, also fulfill~\eqref{eq:factPure} but with concurrence evaluated for the resulting \emph{unnormalized} states. Alternatively, when using the normalized versions, the right side needs to be rescaled by the respective probabilities of $\ket{\chi}$ and $\ket{\phi^+}$ to pass $\Lambda$:
\begin{multline}
C\left( \frac{(\one\otimes\Lambda) \proj{\chi} }{ \Tr[(\one\otimes\Lambda) \proj{\chi}]} \right)
=
C(\ket{\chi})\;\frac{ p_\phi }{ p_\chi }\times \\
\times  C\left( \frac{(\one\otimes\Lambda) \proj{\phi^+} }{ \Tr[(\one\otimes\Lambda) \proj{\phi^+}] } \right),
\end{multline}
where $p_\chi=\Tr[(\one\otimes\Lambda) \proj{\chi}]$ and similarly for $p_\phi$ with $\ket{\phi^+}$ instead of $\ket{\chi}$.

Since pure states are an idealization, let us now consider the case of two two-level systems in an initially mixed state~$\rho_0$. After one of the systems is subjected to non-trivial dynamics $\Lambda$, the state reads
\beq
\rho = (\one\otimes\Lambda) \rho_0.
\eeq
In order to obtain a relation similar to \eqref{eq:factPure}, but for an initially mixed state, we relate the latter with a decomposition into pure states and then employ \eqref{eq:factPure}. For this purpose we use an optimal decomposition of $\rho_0$, namely one for which $C(\rho_0)=\sum_i p_i C(\ket{\chi_i})$. Since the map $\Lambda$ is linear, the mixture of the $\ket{\chi_i}$ individually undergoing the map constitutes the final state $\rho$. However, concurrence is a convex function as every valid entanglement measure, meaning that probabilistic mixing of states cannot increase the entanglement on average, formally
\beq
C \Big[ \sum_i p_i (\one\otimes\Lambda)\proj{\chi_i} \Big]
\leq
\sum_i p_i C[(\one\otimes\Lambda)\proj{\chi_i}].
\eeq
Therefore, we obtain an upper bound when applying \eqref{eq:factPure} to each of the terms on the inequality's right hand side. After collecting the sum over $C(\ket{\chi_i})$ into the initially mixed state's concurrence, we arrive at the evolution equation for mixed state entanglement
\beq \label{eq:factMixed}
C\big[ (\one\otimes\Lambda) \rho_0 \big] \leq C(\rho_0)\; C\big[ (\one\otimes\Lambda) \proj{\phi^+} \big].
\eeq
With this relation we can straightforwardly approach more elaborated  setups where \emph{both} subsystems undergo arbitrary open system dynamics:
\beq
\rho = (\Lambda_1\otimes\Lambda_2) \rho_0.
\eeq
We group the effective result of just one of the maps onto $\rho_0$ into an intermediate state, say $\rho = (\one\otimes\Lambda_2) \rho_1$ with $\rho_1=(\Lambda_1\otimes\one) \rho_0$, and apply the relation for mixed states \eqref{eq:factMixed} twice to yield
\begin{multline} \label{eq:factTwoSides}
C\big[ (\Lambda_1\otimes\Lambda_2) \rho_0 \big] \leq C(\rho_0) \; C\big[ (\one\otimes\Lambda_1) \proj{\phi^+} \big] \times\\
\times C\big[ (\Lambda_2\otimes\one) \proj{\phi^+} \big].
\end{multline}
This relation describes the evolution of entanglement as both subsystems undergo arbitrary dynamics for general initial states. Once more, the behaviour of the maximally entangled state undergoing the local dynamics separately determines the upper bound. Also in this more sophisticated setup the initial condition $C(\rho_0)$ and the dynamics completely separate such that statements about the dynamics, here the evolution of the maximal possible entanglement content in the state, do not require the knowledge of the initial state.

Despite the relation being an inequality in general, the equality holds in many relevant cases, for example when one subsystem of an initially pure state undergoes a series of concatenated dissipative dynamics.

Relations \eqref{eq:factPure} and \eqref{eq:factTwoSides} fostered experimental work using linear optics with entangled photons from spontaneous parametric down-conversion~\cite{Farias:2009,Xu:2009}. In contrast to upper bounds \eqref{eq:factMixed} and \eqref{eq:factTwoSides}, when considering entanglement as a resource, it is more relevant to know how much of it is at least available, i.e.\ guaranteed to be left after a corrupting environment interaction. Such lower bounds are presented by several proposals~\cite{Yu:2008,LiZ:2009,Liu:2009}. 

\begin{figure*}
	\centering
	\includegraphics[width=0.5\linewidth]{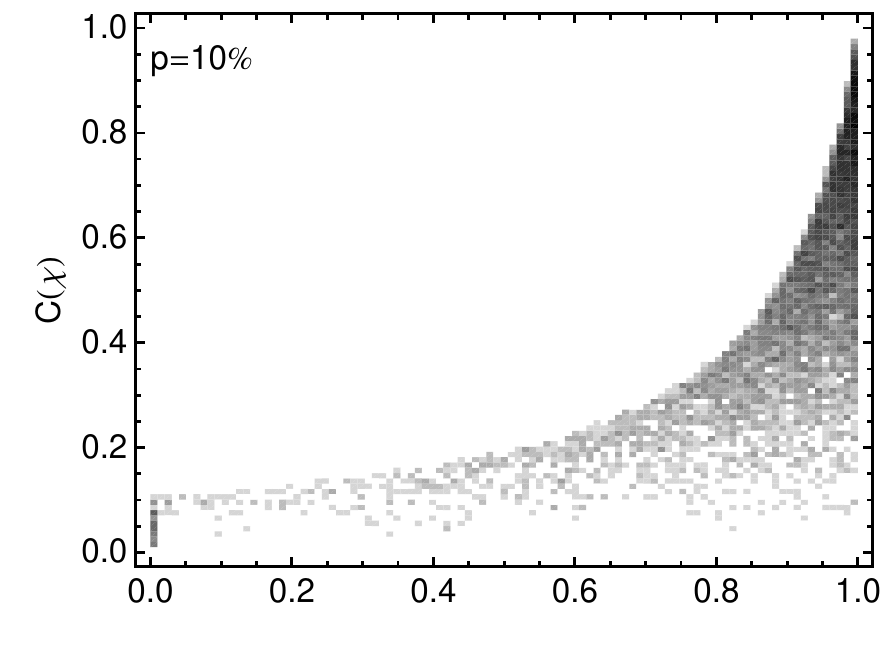}%
	\includegraphics[width=0.5\linewidth]{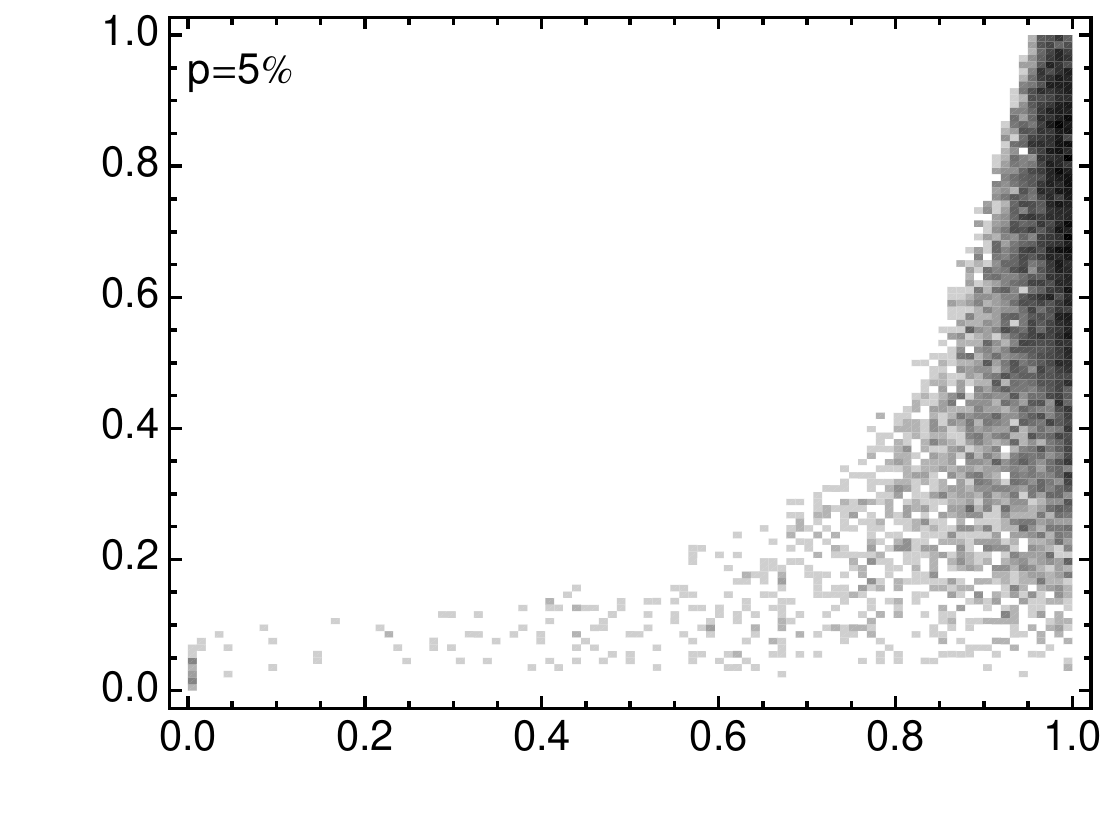}\\
	\includegraphics[width=0.5\linewidth]{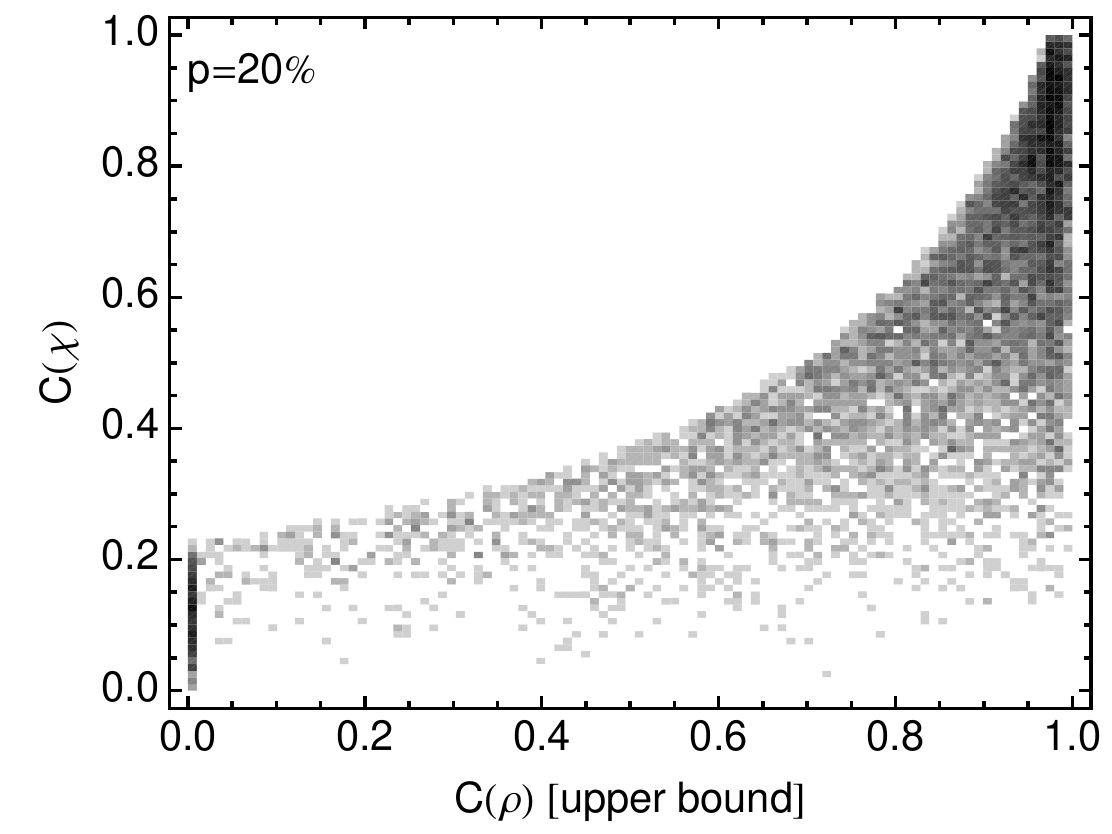}%
	\includegraphics[width=0.5\linewidth]{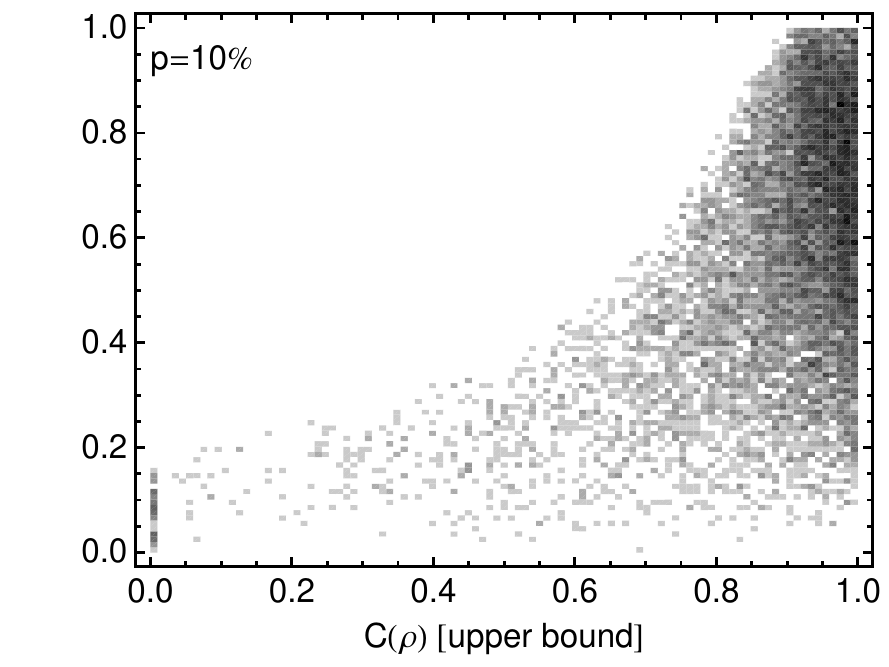}
	\caption{
	Bound tightness for different initial entanglement. 10\,000  initially pure states of a $2\times2$ system are uniformly sampled, and each subsystem evolves under identical incoherent dynamics. The left panels show the data for a phase damping environment in which the subsystems do not exchange excitations with the environment, but lose the coherences in the respective subsystems with probability $p$ due to elastic scattering. The corresponding Kraus operators are $K_0=\sqrt{1-p}\, \one$, $K_1=\sqrt{p}\proj{g}$, and $K_2=\sqrt{p}\proj{e}$, where $\ket{g}$ and $\ket{e}$ denote ground and excited state, respectively. The right panels show data for a dissipative map where $p$ is the probability for an excitation to be released. The corresponding Kraus operators are $K_0=\proj{g}+\sqrt{1-p}\proj{e}$ and $K_1=\sqrt{p}\ket{g}\bra{e}$. Points indicate the concurrence of the final state $C(\rho)$ in units of the upper bound (which is tight for these examples of dynamics) versus the initial amount of concurrence $C(\ket{\chi})$. Darker regions contain more datapoints. }
	\label{fig:points}
\end{figure*}

\begin{figure*}
	\centering
	\includegraphics[width=0.5\linewidth]{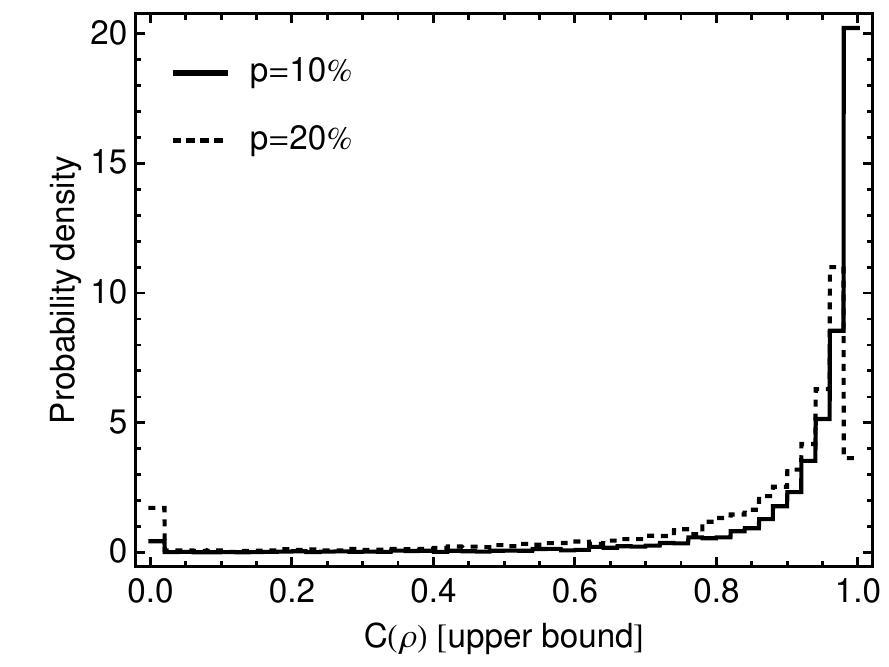}%
	\includegraphics[width=0.5\linewidth]{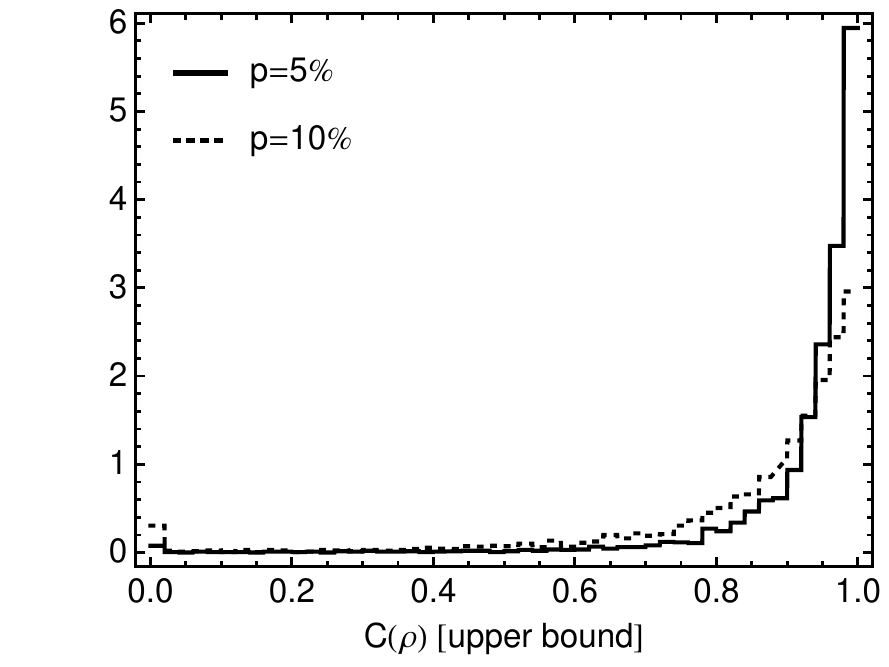}
	\caption{Histograms of the final states' concurrence after a phase damping (left) and dissipation (right) map on both subsystems, respectively. The concurrence after the evolution is plotted in units of the upper bound. All histograms are normalized such that they approximate the probability density and correspond to summing events in Fig.~\ref{fig:points} along columns. The initial states are uniformly sampled. The sample size is 10\,000 states.}
	\label{fig:hist}
\end{figure*}

Although relations \eqref{eq:factMixed} and \eqref{eq:factTwoSides} provide strictly speaking only upper bounds, we demonstrate here that nevertheless they provide a good approximation for experimentally relevant domains. Experiments that aim to create, manipulate, and possibly utilize entangled states usually operate with states of a high purity~\cite{highFid}. Also, the time scales of these experiments fit the time domains where entanglement is mostly preserved. In order to profile decoherence effects during relevant, i.e.\ short time scales, we only need to focus on initially pure states and slight, but non-vanishing, decoherence effects. Since all ingredients in the derivation of \eqref{eq:factTwoSides} are continuous with respect to small changes in the initial quantum state, equality in \eqref{eq:factTwoSides} holds in the limit of pure initial states and vanishing effect of at least one of the maps $\Lambda_{1/2}$. For  small amounts of decoherence acting on both subsystems individually, Fig.~\ref{fig:points} shows (for several examples of maps) that in particular for highly entangled states the upper bound gives a very good approximation. Since the obtained relations for the evolution of concurrence are independent of the particular initial state, also a numerical sampling must avoid any bias. We thus sample uniformly over pure states~\cite{Wootters:1990}. Under such uniform distribution the point density in Fig.~\ref{fig:points} indicates that most states have a decay similar to the bound. This is quantified by the histograms in Fig.~\ref{fig:hist}, which show that for a weak environment influence the great majority of all uniformly sampled states evolves very close to the bound. The upper bound thus leads to a good approximation for most pure states in the considered situations.

\section{Entanglement evolution of high dimensional systems}
\label{lawG}

An extension of the relations \eqref{eq:factPure} and \eqref{eq:factTwoSides} to quantum systems of higher dimension is possible~\cite{Tiersch:2008}. However, the arising complexity in higher dimensions demands its toll. Although concurrence \eqref{eq:concurrencePure} is available for systems of say $d\times d$ dimensions, it can no longer be calculated via the determinant of the state's coefficient matrix \eqref{eq:concDet}. An entanglement quantifier that can be evaluated for mixed states via such a determinant is G-concurrence~\cite{Gour:2005}, which reduces to concurrence for $2\times 2$ dimensions. 
In particular, G-concurrence quantifies the entanglement which requires \emph{all}  the systems dimensions, and thus vanishes for states that live on a strict subspace. For example, a Bell-like state $(\ket{1}\ket{1}+\ket{2}\ket{2})/\sqrt{2}$ in a $3\times 3$ system gives a vanishing G-concurrence despite being entangled. However, among all pure states in $d\times d$ dimensions, those with vanishing G-concurrence are of volume zero. For a pure state $\ket{\chi}=\sum_{i,j=1}^d \chi_{ij} \ket{i}\ket{j}$ G-concurrence is given by
\beq
C_d (\ket{\chi}) = d \abs{\det \chi}^{2/d}
\eeq
and it is extended to mixed states by a similar convex-roof optimization as concurrence in \eqref{eq:convexRoof}. Unfortunately, despite the simple algebraic form for pure states, an analytical expression for mixed states is so far not available. A similar approach as the one for concurrence, however, yields upper and lower bounds~\cite{Gour:2005a}.

Since the argumentation for the evolution equations for entanglement of a $2\times 2$ system does not depend on the dimension but rather on the determinant structure of concurrence, we can straightforwardly apply the same reasoning for G-concurrence and obtain an evolution equation of the very same structure:
\beq
C_d\big[ (\one\otimes\Lambda) \proj{\chi} \big] = C_d(\ket{\chi})\; C_d\big[ (\one\otimes\Lambda) \proj{\phi_d} \big],
\eeq
where the maximally entangled state in $d\times d$ dimensions is $\ket{\phi_d}=\sum_{i=1}^d \ket{i}\ket{i}/\sqrt{d}$. When both subsystems undergo open system dynamics and/or the initial state is mixed an upper bound results \begin{multline}
C_d\big[ (\Lambda_1\otimes\Lambda_2) \rho_0 \big] \leq C_d(\rho_0) \; C_d\big[ (\one\otimes\Lambda_1) \proj{\phi_d} \big] \times\\
\times C_d\big[ (\Lambda_2\otimes\one) \proj{\phi_d} \big].
\end{multline}
Although concurrence and G-concurrence can assume values largely independent of one another, concurrence cannot become arbitrarily large without entanglement that involves all levels of both subsystems and hence G-concurrence becoming non-vanishing. With a vanishing G-concurrence, concurrence at most attains $\sqrt{1-1/(d-1)^2}$ of its maximal value. Therefore, if concurrence exceeds this fraction of its maximal value, G-concurrence is necessarily present.

\section{Conclusion}

Despite of its hard characterization, here we explored  the possibility to design deterministic equations of motion for  entanglement inscribed into two finite dimensional systems. For the case where the initial state is pure, and only one of the subsystems undergoes an incoherent process, the equation of motion for entanglement assumes the form of a simple factorization law -- the first term contains only information about how the entanglement is affected by the dynamics, and the second term scales the first by the  initial amount of entanglement. For more realistic scenarios, where both parts are influenced by an environment and/or the initial state is not pure, the intricate nature of entanglement only allows for the derivation of generic upper bounds. Nevertheless, for most of the situations of experimental interest, the bound represents a fiducial description of the entanglement dynamics. An apparent reduction of complexity in determining the entanglement of a state is obtained when the dynamical process is taken into account. 

It is important to note that, the results here presented are rooted in the Choi-Jamio\l{}kowski isomorphism~\cite{Choi:1975,Jamiolkowski:1972}, and on the mathematical characteristics of the entanglement measures used. For a  more general framework, where the dynamics are not restricted to local ones,  the entanglement measures do not present favorable mathematical structure, and the number of subsystems is not limited to two -- and therefore no dual picture is present -- the quest for a deterministic equation of motion for entanglement seems hopeless. For these cases a statistical approach is more encouraging~\cite{concentration}.

\section{Acknowledgements}
We are grateful for the support of our collaboration by NRF grant 69436 and BMBF grant SUA~08/008.
F. de M. also  acknowledges the support by the Alexander von Humboldt Foundation.



\end{document}